\documentclass[a4paper,11pt]{revtex4}
\usepackage{amssymb}
\usepackage{amsmath}
\usepackage{graphics}
\usepackage{color}
\usepackage[T1]{fontenc}
\usepackage{times,mathptm}
\usepackage[cp1250]{inputenc}
\usepackage{epsfig,epsf}
\newcommand{\be}{\begin{equation}}
\newcommand{\ee}{\end{equation}}

\newcommand\beq{\begin{eqnarray}}
\newcommand\eeq{\end{eqnarray}}
\newcommand\nvec[1]{\textbf{\emph{#1}}}

\begin{document}

%\textbf{Version}: $\beta$-23.12.2013

%$$ \nvec{k} $$

\title{On effects of multiple gluons in $J/\psi$  hadroproduction\footnote{Support of the Polish National Science Centre grant no. DEC-2011/01/B/ST2/03643 is gratefully acknowledged}}

\author{L.\ Motyka}
\email{leszek.motyka@uj.edu.pl}
\author{M.\ Sadzikowski\footnote{Speaker on the International Workshop on Diffraction in High-Energy Physics,
Primošten, Croatia, 2014.}}
\email{mariusz.sadzikowski@uj.edu.pl}
% \email{leszek.motyka@desy.de}
\affiliation{Institute of Physics, Jagiellonian University, Reymonta 4, 30-059
Krak\'{o}w, Poland
}

\begin{abstract}
The three-gluon contribution to $J/\psi$ hadroproduction is calculated within perturbative QCD in the $k_T$ factorization framework.
This mechanism involves double gluon density and enters at a non-leading twist, but it is enhanced at large energies due to large double gluon density at small $x$.  We obtain results for differential $p_T$-dependent cross-sections for all $J/\psi$ polarisations. The rescattering contribution is found to provide a significant correction to the standard leading twist cross-section at the energies of the Tevatron or the LHC at moderate $p_T$. We also discuss
a possible contribution of the rescattering correction to the anti-shadowing effect for $J/\psi$ production in proton - nucleus collisions.
\end{abstract}

\maketitle

\section{Introduction}
Description of $J/\psi$ hadroproduction posted a challenge for theorists since its inception. The Tevatron data showed huge excess of the measured cross-section over the simplest LO QCD predictions, reaching almost two orders of magnitude at large $J/\psi$ transverse momenta, the place that one would expect perturbation approach operate effectively. Since it became clear that the collinear color singlet approximation is insufficient, various theoretical approaches to quarkonium production has been proposed, which try to reconcile theoretical description with experimental data. Among the most prominent is so called color octet mechanism (COM), which allows production of $Q\bar{Q}$ pair (that hadronizes into the meson) also in a color octet state \cite{COM}. This can be combined with different approaches based on collinear factorization (recent papers \cite{collinear_fact1,collinear_fact2}), high energy factorization \cite{kT_fact1,kT_fact2}, color glass condensate \cite{CGC}. There are also other possibilities including rescattering which will be discussed in this letter.

\section{Color octet mechanism and factorization schemes}

Currently, the most successfull model for heavy quarkonia production is based on, so called, color-octet mechanism rooted in non-relativistic  approximation of QCD. In the color octet mechanism one assumes that the $Q\bar{Q}$ pair may be produced not only in the singlet state, but also in the color octet representation. It relies on factorization formula which states that the inclusive cross section for quarkonium production in the hadronic collision can be expressed as the product of inclusive perturbative QCD (pQCD) cross section for producing $Q\bar{Q}$ pair multiplied by non-relativistic QCD matrix elements. The pQCD cross sections can be expanded in powers of the strong coupling constant at the large scale of the order of the heavy quark mass, convolved with parton distributions in collinear version and unintegrated parton distribution in $k_T$-factorization approach. What is crucial for effective predictivity, one assumes existence of universal transition amplitudes from the $Q\bar{Q}$ pair with given quantum numbers to the mesons. The COM has been developed up to the next-to-leading order (NLO) within collinear approximation and it was shown to provide a good global fit of prompt quarkonium production data \cite{collinear_fact1} except of the quarkonium polarization \cite{collinear_fact2}. In $k_T$-factorization framework the emerging picture is quite encouraging, but yet not fully clear. A recent description of the LHC data within the $k_T$-factorization color singlet model approach was quite successful, including even meson polarization \cite{kT_fact3}, but the older description of the Tevatron data required also the color octet contributions \cite{kT_fact2}. Therefore, in this moment the $k_T$-factorization approach also suffers from some deficiencies in providing the consistent global picture of prompt quarkonia production.

\begin{figure}
\centerline{
\includegraphics[width=0.4\columnwidth]{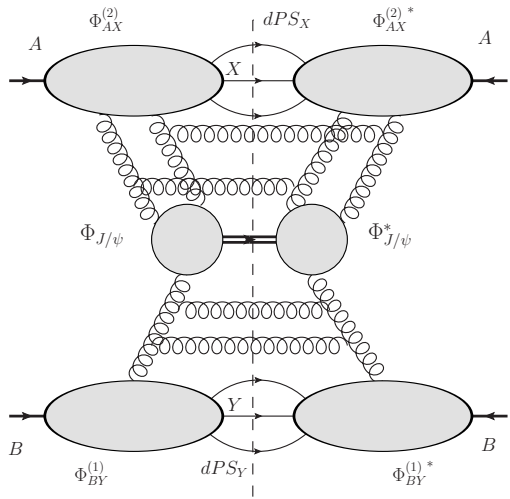}%\hspace{5em}
}
\caption{Relevant diagram for cross section. Similar but flipped diagram also gives contribution.
\label{fig4}
}
\end{figure}

\section{Rescattering correction}

The existing approaches to vector quarkonia hadroproduction, although quite successful, do not provide fully satisfactory global description of data. Thus it may be still necessary to include yet another production mechanism. One of the potentially important corrections may be driven by multiple scattering effects. At the lowest order the correction is given by a partonic amplitude of three-gluon fusion $g+2g\rightarrow Q\bar{Q}$ \cite{Khoze}. The correction employs the double gluon distribution and enters beyond the leading twist. This implies a power suppression w.r.t.\ the standard, two-gluon cross-sections, however, the double gluon distribution at small gluon~$x$ provides a large enhancement factor, that may well reach about 20 in the relevant kinematic domain. We performed a detailed estimate of this rescattering correction for the Tevatron and the LHC energies beyond the collinear, leading logarithmic accuracy of \cite{Khoze}.

The calculations were performed within the $k_T$-factorization approach. The relevant diagram for an inclusive process is shown in Fig.\ 1 with four and two gluons in the $t$-channel. Diagrams with three gluons in the $t$-channel may be neglected as the three gluon state evolution is known to have no energy enhancement. After performing the phase space integrals over remnants $X$ and $Y$ in this contribution one recovers double and single unintegrated gluon distributions originating from $A$ and $B$ respectively. The three gluon-fusion amplitude into the meson
is described in terms of a known impact factor \cite{Bzdak}, with $J/\psi$ polarization vectors in the helicity frame. The proper normalization
of the amplitudes is obtained by matching to collinear cross-sections of single and double parton scattering. We assume factorization of the double gluon density and factorization of the impact parameter dependence. With these assumptions the final formula for the triple gluon contribution to $J/\psi$ production takes the form:
\beq
\label{cross-section2}
&& \frac{d^3\sigma_{pp\rightarrow J/\psi X}}{d\ln\beta d\nvec{p}^2} = \mathcal{N}\frac{R_{\mbox{sh}}^2}{\sigma_{\mathrm{eff}}}\alpha_s^3
\int d^2\nvec{k}d^2\nvec{k}_{1}
\frac{f(\beta,(\nvec{p}_{ }-\nvec{k}_{ })^2)
f(\alpha,k_1 ^2)
f(\alpha,(\nvec{k}_{ }-\nvec{k}_{1 })^2)}
{\left((\nvec{p} -\nvec{k}_{ })^2\nvec{k}_{1 }^{\,2}(\nvec{k}-\nvec{k}_{1 })^2\right)^2} \nonumber\\
&&\left|\Phi_{J/\psi}(\alpha,\beta;\nvec{k}_{1 },\nvec{k}-\nvec{k}_{1 },\nvec{p}_{ }-\nvec{k}_{ };\epsilon)\right|^2
 + (\alpha\leftrightarrow\beta,p_A\leftrightarrow p_B) .
\eeq
where $\alpha$ and $\beta$ are meson longitudinal momentum fractions of the projectiles, rapidity $Y =1/2\log(\alpha/\beta )$,
$\nvec{p}$ is the meson transverse momentum, $\nvec{k},\nvec{k}_1$ gluons momenta, $\epsilon$ is the meson polarization and we combine numerical constants, color factors etc.\ into a normalization constant $\mathcal{N}$. Note the emergence of multiple scattering parameter $\sigma_{\mathrm{eff}}$ related to probability of double scattering and Shuvaev factor $R_{\mathrm{sh}}$~\cite{shuvaev} because off-diagonal gluon densities are involved in the process.

\section{Results}

In the numerical evaluations unintegrated gluon densities were used derived from the CT10 collinear gluon density \cite{CT10} using Kimber-Martin-Ryskin approach \cite{KMR} with the hard scale given by the transverse meson mass. The running strong coupling constants of a gluon with virtuality $\nvec{k}^2$ was evaluated at the scale $m^2 = m_c^2 +\nvec{k}^2$, with $m_c = M_{J/\psi}/2$. We set the multiple
scattering parameter value $\sigma_\mathrm{eff} = 15$~mb, in accordance with Refs.\ \cite{sigma_eff}. In Fig.\ 2a the results of numerical evaluation of the triple gluon correction are shown for the Tevatron energy ($\sqrt{s} = 1.96$~GeV). For a reference, we show in Fig.\ 2b a plot taken from Ref.\ \cite{collinear_fact1}, in which results of CSM and of COM fits are presented at LO and NLO accuracy together with CDF data \cite{CDF}. The comparison of Fig.\ 2a and Fig.\ 2b reveals that the triple-gluon contribution is larger than the CSM collinear contribution in the whole range of transverse momentum and it reaches about 20 -- 25$\%$ of the measured cross-section at low transverse momenta. At larger transverse momenta the relative importance of this correction diminishes. The results for the LHC exhibit very similar pattern, as the results for the Tevatron, so we do not show them here, leaving the broader and more detailed presentation to the forthcoming paper~\cite{motyka_sad}.

\begin{figure}
 \begin{center}
 \includegraphics[width=0.4\columnwidth]{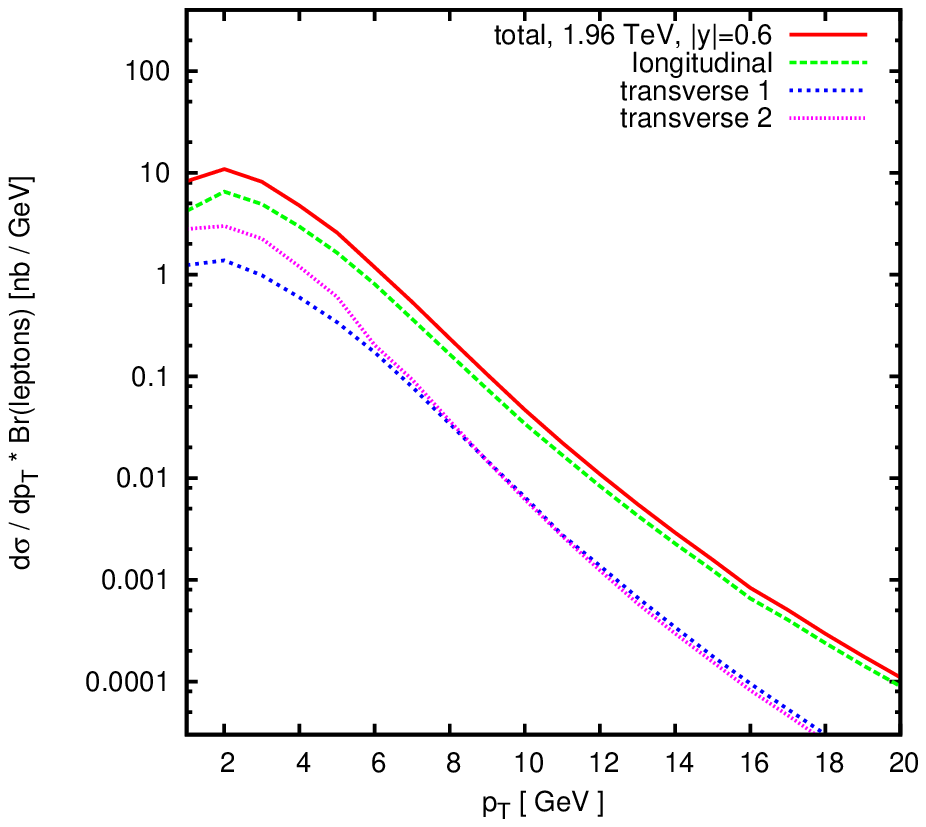}\includegraphics[width=0.4\columnwidth]{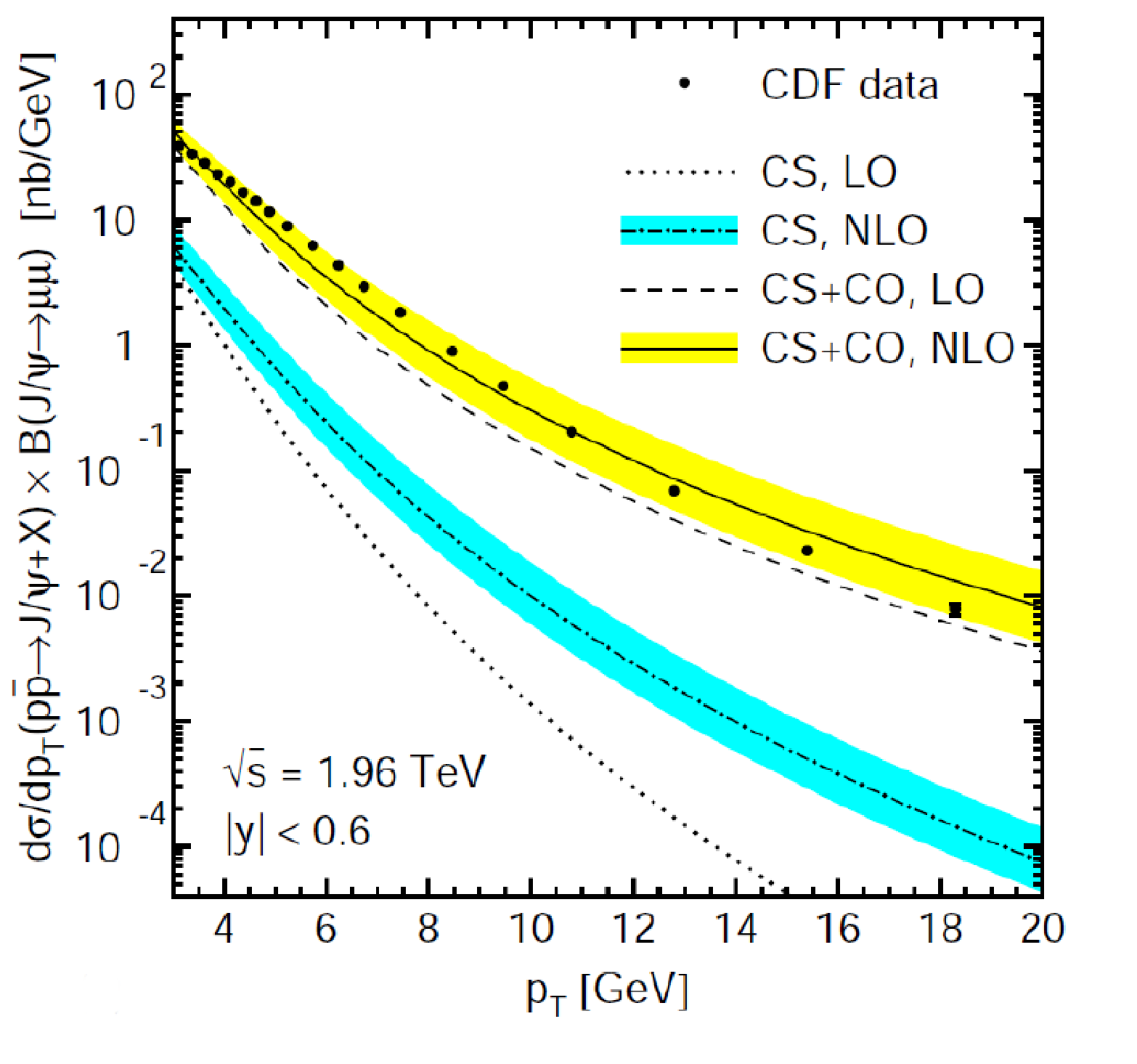}
 \caption{Differential cross section $d\sigma_{pp\rightarrow J/\psi X}/dp_T\times$ BR$(J/\psi\rightarrow \mu^+\mu^-)|_{|Y|<0.6}$
 at the Tevatron $\sqrt{s}=1.96$ TeV: a) rescattering correction and its polarization decomposition; b) the reference:
plot taken from \cite{collinear_fact1}: CDF data \cite{CDF} and LO / NLO fits of the CSM and COM \cite{collinear_fact1}.}
   \label{twist2T}
 \end{center}
\end{figure}

Rescattering process described in this letter leads to a prediction of a nonlinear anti-shadowing effect in the proton-nucleus collision. Consider the process in the proton rest frame. Then, the per-nucleon cross-section of the triple-gluon contribution to $J/\psi$ production is enhanced when 2~gluons from the nucleus interact with a single gluon from the proton. The correction from two-gluons taken from the nucleus is predicted to be large in the proton fragmentation region and in this region one needs to resum also contributions from more gluons. In the nucleus fragmentation region, however, the nuclear gluon densities are moderate and the
triple gluon contribution approximates well the rescattering correction. Thus, in the nucleus fragmentation region we predict that the nuclear modification factor of the rescattering correction, $R^{3g}=\sigma^{3g} _{pA\rightarrow J/\psi X}/A\sigma^{3g}_{pp\rightarrow J/\psi X}$ grows with atomic number $A^{1/3}$. This happens because $A^2$ enhancement of the cross section $\sigma^{3g}_{pA\rightarrow J/\psi X}$ due to the number of nucleons, divided by the effective multiple scattering parameter $\sigma_\mathrm{eff}$ which scales as a nucleus radius $R_A^2\sim A^{2/3}$, which together gives $A^{4/3}$. Numerical estimates for proton-lead collisions at $\sqrt{s} = 5$~TeV per nucleon pair yield an about 10\% enhacement of the per-nucleon cross-section of $J/\psi$ production in the nucleus fragmentation region, which is consistent with the recent data \cite{Alice}. Such non-linear nuclear effects are
particularly useful for a direct experimental determination of the rescattering correction discussed in this paper.

\textbf{Acknowledgement}

This research was supported by the Polish National Science Center grant no. DEC-2011/01/B/ST2/03643.

\end{document}